\documentclass[article,twocolumn,superscriptaddress,prb,showpacs]{revtex4}  
\usepackage{graphicx,amsmath,mathrsfs}
\usepackage{amssymb}
\usepackage{amsthm,multirow}
\usepackage{bm,bbm}
\usepackage{subfigure,xcolor} 
\usepackage{color}
\mathchardef\Gamma="7100
\gdef\bs#1{\boldsymbol{#1}}

\def\ns{^{\vphantom{*}}}
\def\yd{^\dagger}

\def\Bk{{\bs{k}}}
\def\Br{{\bs{r}}}

\def\BB{{\bs{B}}}
\def\BE{{\bs{E}}}
\def\Ba{{\bs{a}}}
\def\Bp{{\bs{p}}}
\def\Bg{{\bs{g}}}
\def\Bj{{\bs{j}}}
\def\Bx{{\bs{x}}}
\def\By{{\bs{y}}}
\def\Bz{{\bs{z}}}
\def\Bpi{{\bs{\pi}}}
\def\xhat{{\hat\Bx}}
\def\yhat{{\hat\By}}
\def\zhat{{\hat\Bz}}
\def\wtH{{\widetilde H}}
\def\CO{{\cal O}}
\def\CE{{\cal E}}
\def\CT{{\cal T}}

\def\oc{{\omega\ns_{\rm c}}}
\def\Vk{V\ns_\Bk}
\def\rhok{\rho\ns_\Bk}
\def\rhob{{\bar\rho}}
\def\rhobk{\rhob\ns_\Bk}

\def\be{\begin{equation}}
\def\ee{\end{equation}}
\def\bea{\begin{eqnarray}}
\def\eea{\end{eqnarray}}

\def\rwk{\Gamma^2_\Bk}

\begin{document}
\title{Non linear conductivity and collective charge excitations in the lowest Landau level}

\author{Assa Auerbach}
\affiliation{Physics Department, Technion, 32000 Haifa, Israel}
\author{Daniel P. Arovas}
\affiliation{Department of Physics, University of California at San Diego, La Jolla, California 92093, USA}

\date{\today }

\begin{abstract}
For weakly disordered fractional quantum Hall phases,  the non linear photoconductivity is related to the 
charge susceptibility of the clean system by a Floquet boost. Thus, it may be possible to probe collective charge modes at finite wavevectors by electrical transport.
Incompressible phases, irradiated at slightly above the magneto-roton gap, are predicted to exhibit  
negative photoconductivity and  zero resistance states with spontaneous internal electric fields.   
Non linear conductivity can probe
composite fermions' charge excitations in compressible filling factors.
\end{abstract}

\maketitle

\begin{figure}[!t]
\begin{center}
\includegraphics[width=7.5cm,angle=0]{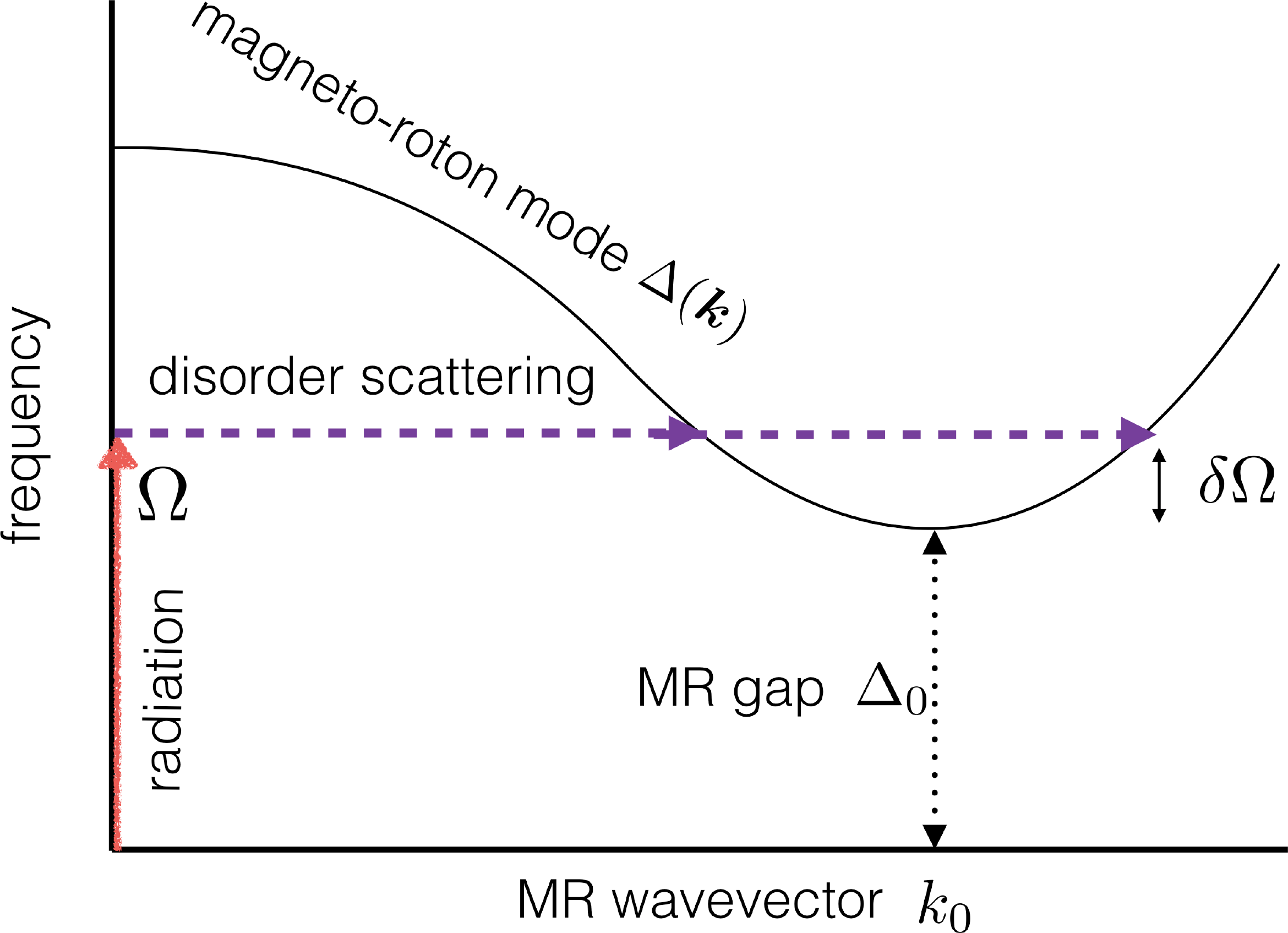}
\caption{Magneto-roton (MR) collective excitation in an incompressible  Laughlin  phase following GMP\cite{GMP1}.
Radiation above threshold frequency $\Omega >\Delta_0$, combined with scattering by disorder yields $\sigma_{xx}>0$,
by Eq.~(\ref{AC}).}
\label{fig:MR}
\end{center}
\end{figure}

Fractional quantum Hall (FQH) phases  exhibit  exotic ground states and many-body
collective modes.  For example, the Magneto-Roton  (MR) mode in Laughlin states was predicted by Girvin, MacDonald and Platzman (GMP)~\cite{GMP1} to have a 
minimal gap $ \Delta\ns_0$ at 
wavevector  $k\ns_0$, (See Fig.\ref{fig:MR}). This mode was seen by  
Raman scattering\cite{pinczuk},  resistivity activation gap with phonon pulses\cite{Phonons}, and microwave absorption with surface acoustic waves\cite{SmetMR}.

Some of these probes are based on the powerful relation between conductivity $\sigma_{xx}(\omega)$ and charge excitations in weakly disordered quantum Hall gases~\cite{Amiram, Fukuyama, GMP-cond}:
\be
\sigma\ns_{xx}(\Omega) = { nc^2 \over\hbar B^2} {1\over A}\sum_\Bk \Gamma^2_\Bk\, k_y^2 \>
{ \chi_0''(\Bk,\Omega)\over \Omega}  +\CO(V^4)\ . 
\label{AC}
\ee
$V(\Br)=A^{-1/2}\sum_\Bk \Vk\,e^{i\Bk\cdot\Br}$
is the random disorder potential taken from a Gaussian ensemble with fluctuation spectrum $\langle V^*_\Bk V\ns_{\Bk'}\rangle = \Gamma^2_\Bk\,
\delta\ns_{\Bk\Bk'}$\,, $\chi_0''(\Bk,\Omega)$ is the  charge susceptibility of the ``clean'' ($V\!=\! 0$), but fully interacting,
system, at wavevector $\Bk$ and frequency $\Omega$.  $B$, $c$, $n$, and $A$ are the perpendicular magnetic field ($\BB=B\zhat$), speed of light, electron density, and system area, respectively.

In this paper we  extend  Eq.~(\ref{AC}) to the {\em non linear} current versus field response. 
We propose new and independent experiments to probe  the collective charge excitations, which would complement 
information given by the linear response probes.   
Our derivation incorporates
strong electric fields into the current reponse function. To go beyond linear Kubo formula, we use a Floquet boost transformation\cite{Ryzhii,Pai}.
Our results retain the full many-body correlations in $\chi_0''$, which make them suitable for the strongly correlated FQH phases.
 
We propose to measure the  DC photoconductivity $\sigma^{\rm photo}_{xx}(\Omega)$ in the presence of an AC radiation field at frequency $\Omega$.
Within GMP theory,   at frequency $\Omega> \Delta_0$,  $\sigma^{\rm photo}_{xx}$ should become {\em singularly large and negative}. 
As a consequence, the system will be  unstable toward a ``zero resistance state'' (ZRS)\cite{Andreev}, where  internal  electric field domains
are spontaneously created~\cite{AYH}.   
ZRS phases, and spontaneous internal fields, were previously observed at weak magnetic fields (high Landau levels)~\cite{Mani,Zudov,SmetNoise}. 
 The photoconductivity  has been calculated by various methods~\cite{Durst,Shi,Dmitriev,Pai}.
At high Landau levels, the dark conductivity is metallic, and negative photocurrent can be induced by strong  microwave  power, which is positively detuned from the cyclotron frequency $\omega_c$.
In contrast, we show that the  ZRS in the lowest Landau level  (LLL) arises from  different phenomena. It involves {\em  intra-Landau Level}  excitations of MR mode which is expected at  $\Omega\ll\omega_c$.
In addition, negative photoconductivity appears already at leading order in radiation power, since the  dark conductivity is zero.

For the compressible phases, such as around $\nu={1\over 2}$,  the dark non linear  current  $\Bj^{\rm dc}_x[E_x]$ is expressed in terms of $\chi''_0$.
Thus, predictions of composite fermions theory for  $\chi_0''(\Bk,\omega)$ could be tested by transport measurements.
We conclude by discussing  experimentally relevant frequency and field scales, and  regime of validity of the weak disorder expansion.

{\em Quantum Hall Hamiltonian} :
Two dimensional electrons in a uniform magnetic field, and a time dependent vector potential $\Ba(t)$, are described by the Hamiltonian
$H=H_0 + V$, where
\begin{equation}
H_0[\Ba] ={1\over 2m}  \sum_{i=1}^N  \big(\Bpi_i - \frac{q}{c}\Ba(t)\big)^2  + \sum_{i<j} \Phi(\Br_i-\Br_j)
\label{Ham}
\end{equation}
and $V=A^{-1/2}  \sum_\Bk \Vk\, \rhok$\,.
$N$ is the number of particles of charge $q$, and  $\Phi$ are the two-body interactions. The density operator is
$\rhok =  \sum_{i=1}^N e^{-i\Bk\cdot\Br_i}$, and the cyclotron momentum for particle $i$ is
$\Bpi_i=\Bp_i-\frac{qB}{2c} \hat{\Bz}\times \Br_i $, obeying $\big[ \pi_i^\alpha,\pi_j^\beta\big]=
i\,{\rm sgn}(q) \, \hbar^2\ell^{-2} \epsilon^{\alpha\beta} \delta_{ij}$, 
where $\ell= \big(\hbar c/ |q| B\big)^{1/2}$ is the Landau length.  For electrons, $q=-e$.
The current density  at time $t$ is
\begin{equation}
\Bj(t)  = {q\over m A } \sum_{i=1}^N \big\langle  U^\dagger (t) \,\Bpi_i \,U(t) \big\rangle   -   {nq^2\over mc} \Ba(t) \ ,
\label{Bjt}
\end{equation}
where  $n= N/A$ , and $\langle  \cdots \rangle  $ denotes thermal and disorder averaging at $t=0$.  $U(t)$ is the time evolution operator.
 
The calculation of the nonlinear photocurrent generalizes the single electron approach of Auerbach and Pai\cite{Pai}, while using the underlying Galilean symmetry of $H_0$.
We proceed in two steps. 

{\em (i) Floquet Boost} :
We decompose $U(t)$ according to
\begin{equation}
U(t)=W(t)\, \CT\, \exp\left\{-{i\over\hbar}\int\limits_0^t \!\!dt'  \>\wtH (t') \right\} \ ,
\end{equation}
where $W(t)\equiv \prod_i e^{-i\Bg(t)\cdot \Bpi_i/\hbar}$ is the unitary Floquet boost.
To cancel  $\Ba(t)$ from $H_0[\Ba]$, $\Bg(t)$ must satisfy
\begin{equation}
{\dot\Bg}+ \oc \zhat\times\Bg(t) = -{\oc\over B}\> \Ba(t) \ .
\label{gOFa}
\end{equation}
where $\oc={eB\over mc}$ and $\Bg(0)=0$. $\Bg(t)$ can be readily solved for any  $\Ba(t)$.

$\wtH =W^\dagger H W - i \hbar\, W^\dagger \dot{W} $ is the Hamiltonian in the boosted frame,
given by $\wtH=\wtH_0+ {\tilde V} + f(t) $, with
\begin{equation}
\begin{split}
 \wtH_0 &= \sum_i  {\Bpi_i^2\over 2m}   +  \sum_{i<j} \Phi (\Br_i-\Br_j)  \\
 {\tilde V}&=  A^{-1/2} \sum_\Bk \Vk\, \rhok \, e^{ i\Bk \cdot   \Bg(t)} \ ,
\label{Htilde}
\end{split}
\end{equation}
and $f(t)$ is an irrelevant $c$-number.

 {\em (ii) Expansion in the disorder potential} :
 The evolution operator is $U(t)=U\ns_0\,\CT\,\exp \Big[-\frac{i}{\hbar}\int\limits_0^t dt'\>V(t')\Big]$,
where $U_0= W(t) \, e^{-i \wtH_0 t/\hbar}$ is the clean evolution operator, and  $V(t)=A^{-1/2}\sum_\Bk \Vk\,\rhok(t)\,e^{i\Bk\cdot\Bg(t)}$,
where $\rhok(t) \equiv e^{i \wtH_0 t/\hbar} \rhok\, e^{-i \wtH_0 t/\hbar}$.  

The evolution of the global Landau  operator $\Pi^\dagger= \sum_i (\Bpi^x_i-i\Bpi^y_i)$ for the disorder-free  Hamiltonian is given by
Kohn's theorem:
\begin{equation}
U_0^\dagger(t) \,\Pi^\dagger\,  U_0(t) =
\exp(i\oc t)\,\Pi\yd - {i\,N\hbar^2\over \ell^2} \,{\rm g}^*(t)\ ,
\label{Pi}
\end{equation}
where ${\rm g}(t)=g\ns_x(t) + i g\ns_y(t)$.
For the clean system, by setting $U\to U_0$ in (\ref{Bjt}),   it is easy to see from (\ref{Pi}) that for a DC field $\BE$, (i) the longitudinal  current 
vanishes at all times, and (ii) the Hall current is the Galilean result  $\Bj = (nec/B) \hat{\Bz} \times \BE$. 

The longitudinal current (\ref{Bjt}) is expanded in powers of the disorder $(V)^n$. The leading order is for $n=2$,  
whose Fourier transform is given by 
\begin{align}
j_x  (\omega) &= {qn\over m \hbar }\>    {1\over A} \sum_\Bk \rwk\> {k_y  \over  \omega + \oc} \label{tj2}
 \\
&\times \int\limits_{-\infty}^\infty \! {d\omega' \over 2\pi}~R_{-\Bk}(\omega -\omega') \,\chi_0''(\Bk,\omega') \, R_{\Bk}(\omega') \ .\nonumber
\end{align}
The {\em electric field factors}   are given by
\be
R_\Bk[\BE(t),\omega)] \equiv   \!\int\limits_{-\infty}^\infty \! \!dt' \>e^{i \Bk\cdot \Bg[\BE]} \>
e^{i  \omega t }\ .
\ee
$\chi_0''$ is the dynamical charge susceptibility of the clean system,
\be
\chi_0''(\Bk,\omega) = {1\over NZ_0} \,{\rm Re}\!  \int\limits_0^\infty \!dt\> e^{i   \omega t } 
  {\rm Tr}\left\{ e^{-  {\wtH_0\over T} }  \big[\rho_{-\Bk}(t), \rho_{\Bk}(0)\big]\right\} \ .
\label{chi}
\ee
Eq.~(\ref{tj2}) applies (to second order in $V$)  for an {\em arbitrary time dependent} electric field.

{\em DC Photocurrent.}
We consider an electric field with two components: a  DC field $E_x \xhat$, and a circularly polarized AC  field $\CE$ of frequency $\Omega$, such that
$\Ba(t) = - c\, \CE\big(\sin(\Omega t)\, \xhat - \cos(\Omega t) \,\yhat\big) -  cE_x  t\, \xhat$.  Solving Eq.(\ref{gOFa}) yields
\begin{align}
 \Bg(t)  &=  {cE_x \over \hbar B\oc} \big( \xhat -\oc t\,\yhat\big) \\
 & \qquad -  {c\,\CE  \over B(\oc -  \Omega) }  \big(  \cos(\Omega t)\, \xhat +  \sin(\Omega t)\, \yhat \big)\ . \nonumber
  \label{Eq-gt}
\end{align}
and
\be
R\ns_\Bk(\omega)=  \sum_{l=-\infty}^{\infty} e^{i\phi\ns_\Bk} J\ns_l (k\lambda)\,\delta\Big( \omega + l \Omega - {ck\ns_y\over B} E_x \Big) 
\label{Rpc}
\ee
where where $k=|\Bk|$ , $J\ns_l(x)$ is a Bessel function of order $l$, and $\lambda=c\,\CE/B\Omega$ measures the
radiation field  strength. $\phi\ns_\Bk$ is an irrelevant phase.

The  nonlinear DC photocurrent as a function of electric fields is
\begin{equation}
\begin{split}
j^{\rm dc}_x(\Omega,\CE,E_x) &= {n c \over B   \hbar} \,{1\over A}\sum_\Bk \rwk \, k_y \!  \sum_{l = -\infty}^{\infty}  
\left|J_l ( k \lambda\ns_\CE)\right|^2  \\
&\qquad\qquad \times  \chi_0'' \left(\Bk\,,  l \Omega + {ck_y \over B}   E_x \right)\ .
\label{NLj}
\end{split}
\end{equation}
In the following, we apply Eq.~(\ref{NLj}) to the incompressible and compressible FQH phases.
  \begin{figure}[!t]
\begin{center}
\includegraphics[width=8.5cm,angle=0]{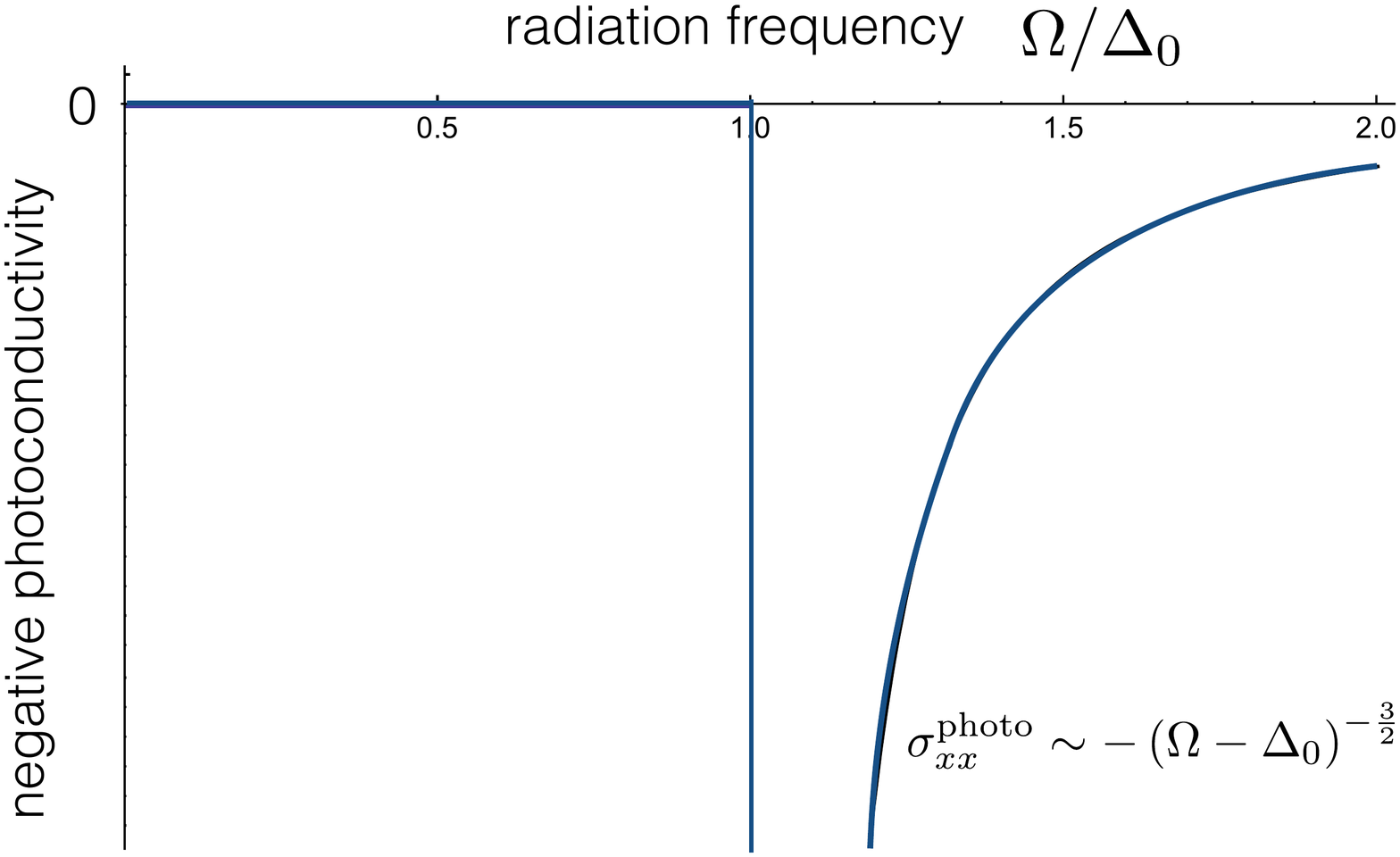}
\caption{Photoconductivity in an incompressible Laughlin phase, given by Eq.(\ref{pcSMA}).
 $\Delta_0$ is the MR gap depicted in Fig.~\ref{fig:MR}. $\Omega$ is the radiation frequency.
At $\Omega>\Delta_0$,  large negative conductivity is predicted.  }
\label{fig:sigma}
\end{center}
\end{figure}

 {\em Incompressible Laughlin states}. 
 At zero temperature,  filling fractions $ nhc/|eB|= \nu_m=1/m $, for odd $m$,  are incompressible.
In the strong magnetic field limit $\oc >> \Delta_0$,  and at frequencies $ \omega\ll \oc$, the  charge susceptibility can be computed {\em within}
the LLL, {\it i.e.\/} using projected density operators $\rho \to \rhobk=P\rhok P$ where $P$ is
the LLL projector. 
The projected structure factor $s(\Bk)= N^{-1}\langle \rhobk\,\rhob\ns_{-\Bk}\rangle$ ,  was computed by GMP\cite{GMP1} using a Monte-Carlo simulation 
of the classical two dimensional one component plasma. 
At low $\Bk$, $S(\Bk) \sim k^4$.  GMP also computed the oscillator strength 
$f(\Bk)= \frac{1}{2}N^{-1}\left\langle \left[ \rhobk, \big[{\bar H},\rhob\ns_{-\Bk}\big]\right]\right\rangle$ for 
the Laughlin state with the LLL-projected Coulomb Hamiltonian.
The  single mode approximation (SMA) {\em assumes} that the spectral weight is exhausted by a single collective mode, i.e.
\be
\chi_0''(\Bk,\omega) \simeq  \pi\, e^{-k^2\ell^2/2} \, s(\Bk) \,\delta\big(\omega - \Delta (\Bk) \big)\ ,
\label{chi-SMA}
\ee
where $\Delta(\Bk)= f(\Bk) /s(\Bk)$ is the
magneto-roton  dispersion.  GMP's calculation is conveniently parametrized by  a Taylor expansion about  $k\ns_0$ 
\begin{equation}
\Delta(\Bk)=  \Delta\ns_0\, \big(1+ a_2  (\delta k\,  \ell)^2+ a_3 (\delta k\, \ell)^3  + \ldots \big) \ .
\end{equation}
 where $\delta k = k-k\ns_0$, and $a_2 >0$.

\begin{figure}[!b]
\begin{center}
\includegraphics[width=8.5cm,angle=0]{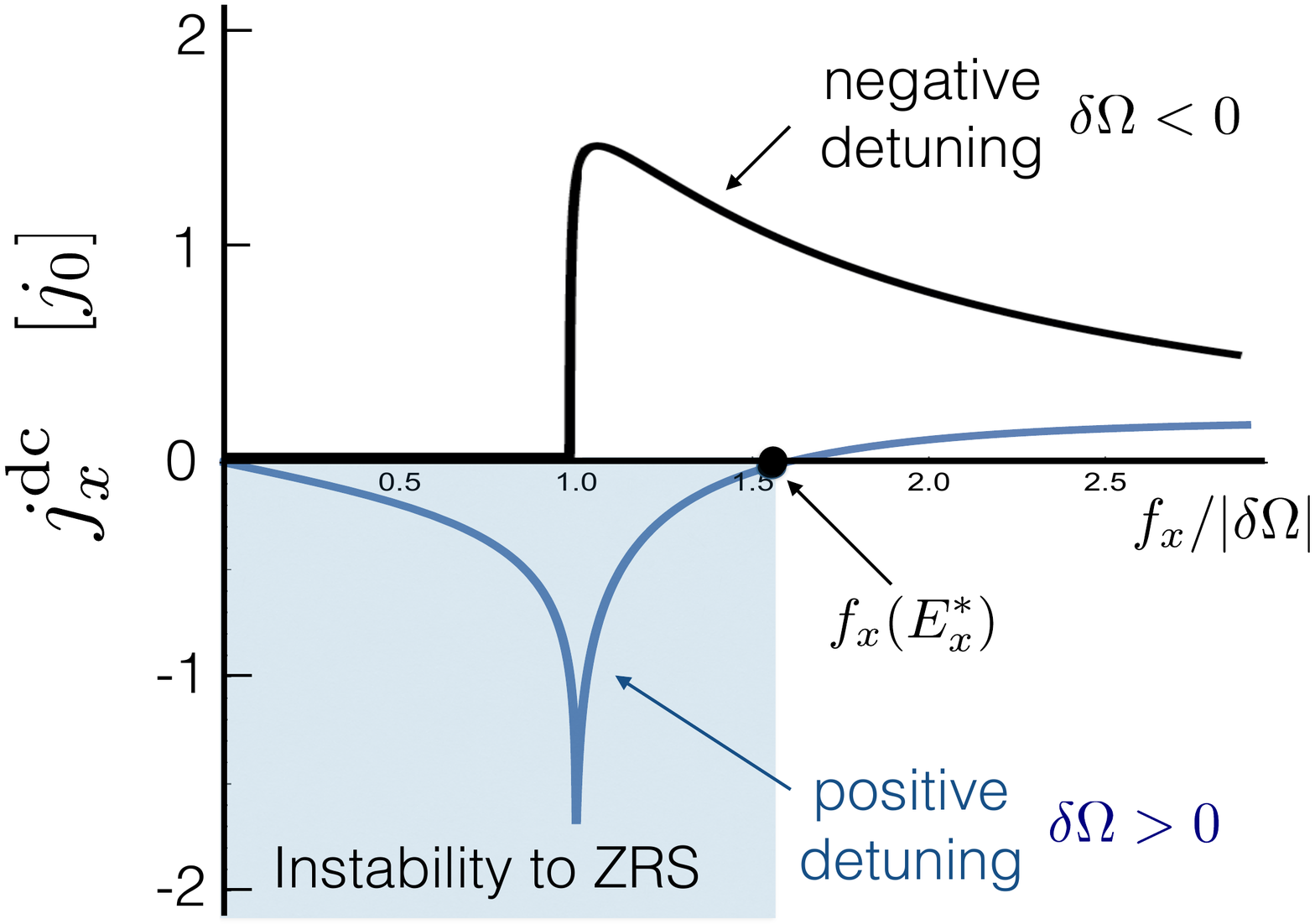}
\caption{Nonlinear photocurrent for an incompressible Laughlin phase,  given by Eq.~(\ref{jdc}).  $f_x(E_x)$ is defined in Eq.~(\ref{fx}). 
For positive detuning $\delta\Omega>0$, (blue curve), the
negative photocurrent is unstable in the shaded region toward a ZRS state with spontaneous  electric field $E_x^*$ given by Eq.~(\ref{Estar}).}
\label{fig:gE}
\end{center}
\end{figure}

Using Eq.~(\ref{chi-SMA}), the photocurrent (at radiation powers $\lambda\ns_\CE/\ell  \ll 1$) is
\begin{align}
j^{\rm dc}_x(\Omega, E_x)   &\simeq   { n c |\lambda\ns_\CE|^2 \over  \hbar  B    } \int\limits_0^\infty \!dk\>  e^{-{1\over 2} (k \ell)^2} k^4 \,\rwk\,s(k)\!\int\limits_0^{2\pi}\! d\theta \sin \theta  \nonumber  \\
&\hskip 0.5in\times 
\delta\! \left(  \Omega -\Delta(k) +{E\ns_x\over B}\,ck \sin \theta \right)\ .  
\label{jE-SMA}
\end{align}
The  {\em detuning frequency}  is defined as
$ \delta\Omega =\Omega- \Delta\ns_0$. The DC electric field defines the electric frequency scale
\be
 f_x= c k\ns_0 E_x /B \ .
 \label{fx}
 \ee
In the regime 
$f_x , |\delta\Omega| \ll \Delta\ns_0$,  the DC photoconductivity (PC) is
\begin{align}
\sigma^{\rm photo}_{xx} & \equiv {d j^{\rm dc}_x(\Omega \over d E_x}\Bigg|_{E_x=0}\\
&\simeq   |\lambda\ns_\CE|^2 \left( { n c^2 \over   \hbar B^2   } \right)\int\limits_0^\infty \!dk\>\rwk\,  k^5\,s(k)  
{d\over d  \Omega} \delta \left(   \Omega -\Delta(k)\right) \nonumber\\
&\simeq - |\lambda\ns_\CE|^2  \left( {n c^2  \Gamma^2_{k\ns_0}  k_0^5 \,s(k\ns_0) \over      4 \hbar  \ell B^2 \sqrt{ a_2\Delta\ns_0 } }\right)
|\delta\Omega|^{-3/2}\>\Theta(\delta\Omega) \ ,\nonumber
\label{pcSMA}
\end{align}
The dependence of $\sigma^{\rm photo}_{xx}$ on radiation frequency is  depicted in Fig.~\ref{fig:sigma}.

At finite driving field $E_x >0$, the photocurrent is
\be
j_x^{\rm dc}  =j_0 ~   h\!\left({  \delta\Omega\over |f_x| }\right) ~{\rm sgn} (f_x)\quad,
\label{jdc}
\ee
where the current scale is 
\be
j_0 \equiv { 2|\lambda\ns_\CE|^2 n \ell  \,k_0^4\,s(k\ns_0)\,\Gamma_{k\ns_0}^2  \over \sqrt{a\ns_2\,\Delta\ns_2} } ~ |\delta\Omega |^{-{1\over 2}}   \quad,
\ee
and $h(u)$ is the universal function,
\begin{equation}
h(u)= |u|^{1/2}\!\!\!\! \int\limits_{(u-1)_+}^{u+1}\!\!\!\!\! {ds\over\sqrt{s}} \> {s-u\over\sqrt{1-(u-s)^2}}\quad,
\end{equation}
where $(u-1)\ns_+=\textsf{max}(0,u-1)$.  Photocurrent {\it versus\/}  electric field at fixed 
detuning frequency is depicted in Fig. \ref{fig:gE}, for both positive and negative detuning frequencies.

{\em ZRS spontaneous field.} 
Negative uniform conductivity signals a thermodynamic instability toward formation of a ZRS state\cite{Andreev} with 
spontaneous internal electric fields.  For a homogeneous quantum Hall phase  these fields are fixed by the minima of the Lyapunov functional
condition\cite{AYH} $ j^{\rm dc}_x(\Omega, \BE^*)=0$. Near the detuning threshold,  we see in Fig.~\ref{fig:gE} that the current vanishes  $f_x^*= 1.6 \,\delta\Omega$,
which by Eq.~(\ref{fx}) yiellds the magnitude of the spontaneous electric fields as,
\be
E^*[{\rm V}/{\rm cm}]= 1.6 \times \delta \Omega  {B\over c k\ns_0} ~\Theta(\delta\Omega) \ .
\label{Estar}
\ee
{\em Notice that the spontaneous field is an independent measure of  the MR wavevector $k_0$.}
For inhomogeneous systems with spatially varying Hall conductivity\cite{Finkler}, there is no Lyapunov functional.   
As a result, spontaneously generated electric fields can fluctuate in time\cite{FH}, 
which may be the source of the experimentally observed telegraph noise\cite{SmetNoise}.

{\em Compressible FQH phases}. At $\nu\simeq {1\over 2}$,   the electronic states are have been described by composite Fermions (CF) \cite{HLR,Jain,CF-exp,Shankar}, which
see an effective weak magnetic field. Their effective ``Fermi energy scale'' $\epsilon^*_F$ is determined  by intra-LLL coulomb interactions.
Read~\cite{Read98}  computed the long wavelength dynamical charge susceptibility to go as
\be
\chi_{CF}''(\Bk,\omega) \sim e^{-k^2\ell^2/2}  {\hbar^2 \omega\over (\epsilon^*_F)^2 \,   (k\ell)^3} 
\label{chi-half}
\ee
Near $\nu=\frac{1}{2}$, Park\cite{Park} has proposed to search for ZRS by photoconductivity above the CF  effective Landau level  spacing.
Since the dark system is metallic,  negative conductivity could be achieved at high radiation power if CF Landau levels are sharp modes at finite wavevectors.

Since the excitations are gapless, we can use the 
nonlinear  current as a probe to the low energy excitations. 
In the  ``dark'' DC case, i.e. $\BE(t)= E_x \,\hat{\Bx}$, Eq.(\ref{gOFa}) yields   $\Bg^{\rm dc}(t)=(cE_x/B\oc)\,(\xhat -\oc t\,\yhat)$.
Thus,  
\be
R_\Bk^{\rm dc}(\omega) =  2\pi \exp\!\bigg({i ck_x E_x \over B\oc}\bigg) \>
\delta \!\left(  \omega - {ck_y \over B } E_x \right) \ ,
\label{Rdc}
\ee
and Eqs.~(\ref{tj2}) and (\ref{Rdc})) yield  the dark nonlinear longitudinal current,
\be
j_x^{\rm dc}   =    {nc  \over \hbar B   } {1\over A} \sum_\Bk  \rwk\> k_y \>
  \chi_0''\!\left( \Bk\,, { ck_y \over B} E_x \right)   ,
\label{NLDark}
\end{equation}
The low field limit yields the linear conductivity   
\be
\sigma_{xx}(0) = {d \over d {E_x}} j_x^{\rm dc}= {nc^2  \over \hbar B^2   } {1\over A} \sum_\Bk  \rwk\> k_y^2 \>
  {d\over d\omega } \, \chi_0''\!\left( \Bk\,, 0 \right) \ ,
  \ee
which coincides with  the DC limit of Eq.~(\ref{AC}).

The  characteristic  wave vectors which dominate $\Gamma_\Bk$  in the LLL are of the order of $\ell^{-1}$.
The  dark DC conductivity at the compressible filling fraction $\nu$ is of the scale
\bea
\sigma_{xx} (0)  \sim    { \nu q^2\over h}     {\Gamma^2  \over (\epsilon^*_F)^2 }  \big( 1+{\cal O}(\Gamma/\epsilon_F) \big) \ .
\label{OOM}
\eea
Away from $\nu={1\over 2}$  the composite fermion theory predicts  resonances in $\chi''_0(\omega)$ corresponding to the spectrum of  ``CF Landau levels''.   
Such resonances should appear as oscillations in  the current by Eq.~(\ref{NLDark}).

 {\em Experiments}.  
Photoconductivity measurements in FQH samples were carried out by some time ago\cite{meiselsPC,HeiblumPC}. Oscillatory magnetoresistance  was reported without
comparison to  theoretical calculations. The published data did not show indications of ZRS effects in the   Laughlin phases. However,  the frequency scale  may have been below the magnetoroton threshold:
For filling fraction  $\nu$=1/3 at carrier density  $n$=7.6$\times$10$^{10}$~cm$^{-2}$,  the magnetic field is $B$=9.5~T, 
and $\hbar \omega_c\sim$~100~K. The  resistivity activation gap\cite{meiselsPC}, which is expected to be similar to $\Delta_0$,  was about  210 GHz,
which was  higher than the microwave frequencies used in these experiments. 
Recent advances in terahertz spectroscopy may open the door to photoconductivity 
measurements in the FQH regime.    It would be instructive to compare e.g. the magnetoroton gap by photoconductivity to the activation gap of resistivity.

For the compressible phases, we propose to measure Eq.~(\ref{NLDark}).
The relation between the charge excitation frequency  $\bar{\omega}_{\rm charge}$ and the corresponding electric field $\bar{E}_x$ is  
\bea
\bar{E}_x &=&  B\ell \bar{\omega}_{\rm charge}/  c \nonumber\\
&=&  0.26~  \bar{\omega}_{\rm charge}[{\rm GHz}] ~ \big(B[{\rm T}]\big)^{3\over 2}~~ [{\rm V/cm}] \ ,
\label{Ethr}
\eea
where we have used  $\ell = 26\,{\rm nm} \sqrt{B[{\rm T}]}$.

{\em Validity of weak disorder expansion}.  Eqs.~(\ref{tj2}), (\ref{NLj}) and (\ref{NLDark}) are the
second order expansion of the current in the disorder potential $V$. 
Note that  these expressions have finite LLL limits for $\oc \to \infty$, keeping $\nu<\infty$.\cite{FDMH}
The weak disorder expansion of the longitudinal current is valid under the following conditions:

1.  $\sigma_{xx}=0$ in the clean limit. Note that this condition fails for an ordinary clean metal at zero field.

2.  Landau level broadening. In the absence of interactions  the Landau level are infinitely degenerate. The degeneracy is lifted by  Coulomb interactions which introduce
the intra-Landau level charge excitation scales $ \bar{\omega} =\Delta_0,\epsilon_F^*/\hbar$ we have seen above.
These energy scales control the higher order corrections by powers of  $\Gamma /\hbar\bar{\omega} \ll1$.

3. Density matrix is at equilibrium.   
It is expected that strong time dependent electric fields  modify the density matrix at long times, and produce strong effects on the DC photocurrent~\cite{Dmitriev}. 
Our analysis above did not take non equilibrium effects on the density  matrix into account, which is justified in cases of {\em rapid thermalization}.
In other words, we assume short non radiative relaxation time  by phonons  $\tau_{\rm inel} \ll  \tau_{\rm tr}$,
where  the transport time $  \tau_{\rm tr} = \sigma_{xx} m / e^2 n$ is  long in the weak disorder limit.   
 
4. By applying the derivation of Eq.~(\ref{NLDark}) to the transverse current $j_y(E_x)$, the  Hall conductivity $\sigma^{\rm dc}_{xy}=nqc/B$ gains no corrections at any finite order in $V$.
Thus,  plateaux of $\sigma_{xy}=\nu e^2/h$ in the incompressible phases are necessarily non perturbative effects in $V$ such as nucleation of  
localized quasiparticles or motion of domain edges in the long range potential landscape.

 {\em Acknowledgements} We thank F. D. M. Haldane, S. Kivelson, G. Murthy, E. Shimshoni and J. Smet for useful discussions. We acknowledge support from the
US-Israel Binational Science Foundation grant 2012233 and  the Israel Science Foundation, and thank the Aspen Center for Physics,  supported by the NSF-PHY-1066293, for its hospitality
\bibliographystyle{unsrt}

\bibliography{NLT}

\begin{thebibliography}{10}

\bibitem{GMP1}
S.~M. Girvin, A.~H. MacDonald, and P.~M. Platzman.
\newblock Magneto-roton theory of collective excitations in the fractional
  quantum hall effect.
\newblock {\em Phys. Rev. B}, 33:2481--2494, Feb 1986.

\bibitem{pinczuk}
A.~Pinczuk, B.~S. Dennis, L.~N. Pfeiffer, and K.~W. West.
\newblock Light scattering by collective excitations in the fractional quantum
  hall regime.
\newblock {\em Physica B: Condensed Matter}, 249:40--43, 1998.

\bibitem{Phonons}
C.~J. Mellor, R.~H. Eyles, J.~E. Digby, A.~J. Kent, K.~A. Benedict, L.~J.
  Challis, M.~Henini, C.~T. Foxon, and J.~J. Harris.
\newblock Phonon absorption at the magnetoroton minimum in the fractional
  quantum hall effect.
\newblock {\em Phys. Rev. Lett.}, 74:2339--2342, Mar 1995.

\bibitem{SmetMR}
I.~V. Kukushkin, J.~H. Smet, V.~W. Scarola, V.~Umansky, and K.~von Klitzing.
\newblock Dispersion of the excitations of fractional quantum hall states.
\newblock {\em Science}, 324(5930):1044--1047, 2009.

\bibitem{Amiram}
A.~Ron and N.~Tzoar.
\newblock Interaction of electromagnetic waves with quantum and classical
  plasmas.
\newblock {\em Phys. Rev.}, 131:12--20, Jul 1963.

\bibitem{Fukuyama}
H.~Fukuyama, Y.~Kuramoto, and P.~M. Platzman.
\newblock Many-body effect on level broadening and cyclotron resonance in
  two-dimensional systems under strong magnetic field.
\newblock {\em Phys. Rev. B}, 19:4980--4985, May 1979.

\bibitem{GMP-cond}
P.~M. Platzman, S.~M. Girvin, and A.~H. MacDonald.
\newblock Conductivity in the fractionally quantized hall effect.
\newblock {\em Phys. Rev. B}, 32:8458--8461, Dec 1985.

\bibitem{Ryzhii}
V.~I. Ryzhii.
\newblock Photoconductivity characteristics in thin films subjected to crossed
  electric and magnetic fields.
\newblock {\em Sov. Phys. Solid State}, 11(9):2078--2080, 1970.

\bibitem{Pai}
A.~Auerbach and G.~V. Pai.
\newblock Nonlinear current of strongly irradiated quantum hall gas.
\newblock {\em Phys. Rev. B}, 76:205318, Nov 2007.

\bibitem{Andreev}
A.~V. Andreev, I.~L. Aleiner, and A.~J. Millis.
\newblock Dynamical symmetry breaking as the origin of the zero-dc-resistance
  state in an ac-driven system.
\newblock {\em Phys. Rev. Lett.}, 91:056803, Aug 2003.

\bibitem{AYH}
A.~Auerbach, I.~Finkler, B.~I. Halperin, and A.~Yacoby.
\newblock Steady states of a microwave-irradiated quantum-hall gas.
\newblock {\em Phys. Rev. Lett.}, 94:196801, May 2005.

\bibitem{Mani}
R.~G. Mani, J.~H. Smet, K.~von Klitzing, V.~Narayanamurti, W.~B. Johnson, and
  V.~Umansky.
\newblock Zero-resistance states induced by electromagnetic-wave excitation in
  gaas/algaas heterostructures.
\newblock {\em Nature}, 420(6916):646--650, 2002.

\bibitem{Zudov}
M.~A. Zudov, R.~R. Du, L.~N. Pfeiffer, and K.~W. West.
\newblock Evidence for a new dissipationless effect in 2d electronic transport.
\newblock {\em Physical review letters}, 90(4):046807, 2003.

\bibitem{SmetNoise}
S.~I. Dorozhkin, L.~Pfeiffer, K.~West, K.~von Klitzing, and J.~H. Smet.
\newblock Random telegraph photosignals in a microwave-exposed two-dimensional
  electron system.
\newblock {\em Nature Physics}, 7(4):336--341, 2011.

\bibitem{Durst}
A.~C. Durst, S.~Sachdev, N.s Read, and S.~M. Girvin.
\newblock Radiation-induced magnetoresistance oscillations in a 2d electron
  gas.
\newblock {\em Physica E: Low-dimensional Systems and Nanostructures},
  20(1):117--122, 2003.

\bibitem{Shi}
J.~Shi and X.~C. Xie.
\newblock Radiation-induced ``zero-resistance state'' and the photon-assisted
  transport.
\newblock {\em Phys. Rev. Lett.}, 91:086801, Aug 2003.

\bibitem{Dmitriev}
I.~A. Dmitriev, M.~G. Vavilov, I.~L. Aleiner, A.~D. Mirlin, and D.~G. Polyakov.
\newblock Theory of microwave-induced oscillations in the magnetoconductivity
  of a two-dimensional electron gas.
\newblock {\em Phys. Rev. B}, 71:115316, Mar 2005.

\bibitem{Finkler}
I.~Finkler, B.~I. Halperin, A.~Auerbach, and A.~Yacoby.
\newblock Domain patterns in the microwave-induced zero-resistance state.
\newblock {\em Journal of statistical physics}, 125(5-6):1093--1107, 2006.

\bibitem{FH}
I.~G. Finkler and B.~I Halperin.
\newblock Microwave-induced zero-resistance states are not necessarily static.
\newblock {\em Physical Review B}, 79(8):085315, 2009.

\bibitem{HLR}
B.~I. Halperin, P.~A. Lee, and N.~Read.
\newblock Theory of the half-filled landau level.
\newblock {\em Physical Review B}, 47(12):7312, 1993.

\bibitem{Jain}
J.~K. Jain.
\newblock {\em Composite fermions}.
\newblock Cambridge University Press, 2007.

\bibitem{CF-exp}
V.~J. Goldman, B.~Su, and J.~K. Jain.
\newblock Detection of composite fermions by magnetic focusing.
\newblock {\em Physical review letters}, 72(13):2065, 1994.

\bibitem{Shankar}
R.~Shankar and G.~Murthy.
\newblock Towards a field theory of fractional quantum hall states.
\newblock {\em Physical review letters}, 79(22):4437, 1997.

\bibitem{Read98}
N.~Read.
\newblock Lowest-landau-level theory of the quantum hall effect: The
  fermi-liquid-like state of bosons at filling factor one.
\newblock {\em Phys. Rev. B}, 58:16262--16290, Dec 1998.

\bibitem{Park}
K.~Park.
\newblock Radiation-induced zero-resistance state at low magnetic fields and
  near half-filling of the lowest landau level.
\newblock {\em Phys. Rev. B}, 69:201301, May 2004.

\bibitem{meiselsPC}
R.~Meisels, F.~Kuchar, J.J. Harris, and C.T. Foxon.
\newblock Microwave response of quasi-particles in the fqhe.
\newblock {\em Surface Science}, 263(1):76 -- 80, 1992.

\bibitem{HeiblumPC}
Y.~Guldner, M.~Voos, J.~P. Vieren, J.~P. Hirtz, and M.~Heiblum.
\newblock Microwave photoresistivity of a two-dimensional electron gas and the
  fractional quantum hall effect.
\newblock {\em Phys. Rev. B}, 36:1266--1268, Jul 1987.

\bibitem{FDMH}
F.D.M. Haldane~Private communication.

\end{thebibliography}

 \end{document}